\documentclass[conference]{IEEEtran}

\usepackage[inline]{enumitem}
\usepackage{amssymb}

\begin{document}

\title{Developing a Spatial-Temporal Contextual and Semantic Trajectory Clustering Framework}

\author{\IEEEauthorblockN{Ivens Portugal}
\IEEEauthorblockA{David R. Cheriton School\\ of Computer Science\\
    University of Waterloo\\
    Waterloo, ON\\
    iportugal@uwaterloo.ca}
\and
\IEEEauthorblockN{Paulo Alencar}
\IEEEauthorblockA{David R. Cheriton School\\ of Computer Science\\
    University of Waterloo\\
    Waterloo, ON\\
    palencar@cs.uwaterloo.ca}
\and
\IEEEauthorblockN{Donald Cowan}
\IEEEauthorblockA{David R. Cheriton School\\ of Computer Science\\
    University of Waterloo\\
    Waterloo, ON\\
    dcowan@csg.uwaterloo.ca}
}

\maketitle

\begin{abstract}
This paper reports on ongoing research investigating more expressive approaches to spatial-temporal trajectory clustering. Spatial-temporal data is increasingly becoming universal as a result of widespread use of GPS and mobile devices, which makes mining and predictive analyses based on trajectories a critical activity in many domains. Trajectory analysis methods based on clustering techniques heavily often rely on a similarity definition to properly provide insights. However, although trajectories are currently described in terms of its two dimensions (space and time), their representation is limited in that it is not expressive enough to capture, in a combined way, the structure of space and time as well as the contextual and semantic trajectory properties. Moreover, the massive amounts of available trajectory data make trajectory mining and analyses very challenging. In this paper, we briefly discuss \begin{enumerate*}[label=(\roman*)] \item an improved trajectory representation that takes into consideration space-time structures, context and semantic properties of trajectories; \item new forms of relations between the dimensions of a pair of trajectories; and \item big data approaches that can be used to develop a novel spatial-temporal clustering framework.\end{enumerate*}
\end{abstract}


%
\IEEEpeerreviewmaketitle

\section{Introduction}
\label{Introduction}

A spatial-temporal trajectory is defined as the recording of the position and timestamp of an object at specific times \cite{Frentzos2009}. The widespread use of the GPS devices (e.g. in cars and mobile phones) has enabled the generation of vast amounts of spatial-temporal data about the movement of people, goods, and objects. Trajectory analysis then emerge as the research area that focuses on manipulating, processing, and analyzing trajectory data to discover valuable hidden information, such as patterns and novel insights. Today, trajectory analysis is used, for example, to recommend new routes to travellers \cite{Cui20171}, or faster ways to deliver goods \cite{Zhu20141089}.

It is important to distinguish the dimensions that make up trajectory data \cite{Sakouhi2018390}. One of these dimensions is the spatial dimension, which relates to the location of objects in the space, such as their latitude and longitude. Another dimension is the temporal dimension, which describes the rate at which the spatial information about the object changes as it moves.

One of the ways to analyze trajectories is using clustering techniques. Trajectory clustering \cite{Mazimpaka201661} is an unsupervised method whose goal is to group trajectories based on their dimensions (spatial, temporal) and reveal patterns, commonalities, or ease predictions. Clustering algorithms rely on the definition of similarity between elements, or its inverse: a distance function \cite{Magdy2016613}. In terms of trajectories, similarities may be described based on their dimensions. Similarities based on the spatial dimension take into consideration the length and the path followed by a trajectory \cite{Yuan2017123}, whereas similarities based on the temporal dimension take into consideration the relationship between the time duration of the trajectories being compared (e.g. whether they overlap) \cite{Kostakis2011}.

Although some similarity definitions have been proposed in the literature, their descriptions still remain limited. The structure space and time as well as contextual and semantic properties can be used to improve the expressiveness of trajectory definitions and generate new similarity metrics for trajectory clustering techniques \cite{Le201258}. Similarity definitions can also be based on trajectory context and semantic of movement \cite{Furtado2016280}, which takes into consideration the surrounding elements and reasons related to the moving object (e.g. number of passengers in a vehicle, or going to work). Therefore, context and semantics can be used as additional trajectory dimensions. Moreover, there is a need for implementation framekworks to process the large amounts of data produced in the trajectory domain \cite{Kelley201441}.

In summary, three problems are identified that guide our ongoing research work. They are the lack of: \begin{enumerate*}[label=(\roman*)] \item similarity metrics that take into account improved representations of trajectory data, including its contextual and semantic dimensions; \item trajectory clustering techniques that benefit from improved similarity relations in each trajectory dimension; \item and big data frameworks that supports trajectory analysis using new, improved, and more expressive similarity functions and clustering techniques \end{enumerate*}.

The overarching goal of the research is to improve trajectory similarity models by \begin{enumerate*} [label=(\roman*)] \item taking into consideration spatial, temporal, contextual, and semantics information; \item creating new ways to enrich clustering techniques; and \item developing a novel supporting framework for new trajectory analysis clustering techniques, applications, and evaluation studies \end{enumerate*}. In this way, Data and Machine Learning scientists will be provided new ways of comparing trajectories, calculating outliers, predictions, and drawing data analysis conclusions using clustering techniques.

This paper is strucutred as the following. Section \ref{Background} shows background research on the areas of trajectory definition, similarity, clustering, and big data support. Section \ref{State of the Art Challenges} discusses the main challenges that this research is addressing. Section \ref{Spatial-Temporal Clustering Framework} describes our proposed approach to solve the problems previously described. Lastly, Section \ref{Conclusion} presents our final comments and conclusions.

\section{Background}
\label{Background}

This section presents current approaches to trajectory description, similarity definition, and big data framework development, which are the main research areas ongoing research is inserted. Each of the areas has limitations, which are discussed in section \ref{State of the Art Challenges}.

\subsection{Trajectories and Spatial-Temporal Data}
Spatial-temporal data relates to two dimensions of application data in applications: the spatial and the temporal dimensions. The data describes the spatial properties of objects as a function of time \cite{Parent20061}. The occurrence of tornadoes in a country or people walking in a mall are examples of spatial-temporal data. More formally, spatial-temporal data describes the location $x$ of an element at a particular time $t$. $$(x, t)$$ where $x \in \mathbb{R}^n$ in a theoretical multidimensional world, but is usually described in terms of latitude, longitude, and altitude (or simply x, y, and z) in the $\mathbb{R}^3$. Some researchers \cite{ester2001algorithms} describe the spatial and the temporal representation of this type of data and the following paragraphs review them.

In the spatial representation, objects may be points, lines, regions, or polygons, depending on the dimension that they are represented (zero, one, two, or three, respectively). Although objects in higher dimensions can be represented, their applicability in the real world are limited. Another representation of spatial-temporal objects is based on their edges (end points) or the points in their interior (such as pixels in a screen). The former is the vector representation while the latter is the raster representation.

The spatial representation of the relationship between objects can be divided into topological, distance, and direction, and measurement relations. Topological relations are usually based on objects' boundaries, interior, or complements. Topological relations are, assuming that \textit{A} and \textit{B} are spatial-temporal objects, \textit{A disjoint B}, \textit{A meets B}, \textit{A overlaps B}, \textit{A equals B}, \textit{A covers B}, \textit{A covered by B}, \textit{A contains B}, \textit{A inside B}. Distance relations consider two points of comparison and calculates the distance between two objects. It heavily depends on the definition of the points of comparison (e.g. edges, centroids). Direction relations assume that an object \textit{A} is in a virtual coordinate system and assess where all, the majority, or a specific point of comparison of object \textit{B} is in relation to object \textit{A}. Examples of direction relations are \textit{A to the north of B}, \textit{A below B}, or \textit{A to the right B}. Measurement relations take into consideration properties of objects (e.g. length, perimeter, distance) to specify comparisons such as \textit{A bigger than B} and \textit{A longer than B}.

The temporal representation of data describes the time in which objects exist. Time may be represented as a point or an interval. Another classification of the time representation deals with the way it is associated with the spatial-temporal objects. Some objects have a location that does not change over time (e.g. occurrence of volcano eruptions). This time representation is referred to as \textit{time snapshots}. \textit{Updated snapshots} are temporal descriptions of objects that move over time, but scientists and experts do not keep a history of this movement, only having access to the  most recent information. Lastly, when the history of an object's movement is being captured and stored for analysis, then this time description is referred to as a \textit{time series}. Some properties of time representations that are considered when representing time are its granularity  (dense or coarse), density (discrete or continuous) or the representation of the lifespan of an object (whether a discrete time representation or by the temporal difference between its start and end times). The relationship between two time representations are described in \cite{Allen1983832} and includes, assuming \textit{X} and \textit{Y} are time intervals with length, such that $0 \leq$ length $< \infty$, \textit{X before Y}, \textit{X equal Y}, \textit{X meets Y}, \textit{X overlaps Y}, \textit{X during Y}, \textit{X starts Y}, \textit{X finishes Y}.

A trajectory is, in simple terms, the sequence of positions that a moving spatial-temporal object has taken during a specific time \cite{Aydin2016133}. A trajectory is modeled as a finite discrete sequence of snapshots alongside the spatial-temporal object's position in space, as in $$trajectory = [(x_0, t_0), (x_1, t_1), (x_2, t_2), ..., (x_m, t_m)] = [x,t]$$ where $x_i \in \mathbb{R}^n$, $t_i \in \mathbb{R}$, $0 \leq t_i < \infty, 0 \leq i \leq m$, and $m$ is the length of the trajectory. Although objects can move in an arbitrary dimension, it is common to limit the number of spatial dimensions to three for real world applications. Moreover, details about the shape of the spatial-temporal object are omitted in this definition, but certainly relevant for some domains \cite{Massaâbi2018343}.

The inclusion of semantic data into trajectories has been investigated \cite{Sakouhi2018390}, with reasonings about the goals of the movement, or the semantic meaning of stops. However, this is an area of ongoing reasearch and more studies about data representation and relationship are required. In addition, trajectory data is volumous and as a consequence rich for analyses. Some approaches described in the literature provide various methods in this domain were provided \cite{Mazimpaka201661}. These methods can be mainly classified into clustering or classification methods. They are usually applied to perform pattern mining, outlier detection, or prediction.

\subsection{Similarity and Clustering}
Trajectory clustering techniques are used to analyze trajectory data, trying to group them based on their similarities or differences in spatial and temporal dimensions \cite{ester2001algorithms}. According to \cite{Yuan2017123}, clustering techniques can be hierarchical, partition-based, density-based, and grid-based. Hierarchical clustering methods start with a basic set of clusters and merges or splits them based on a stopping criteria. The result is a tree of clusters, called a \textit{dendogram}. A popular hierarchical clustering algorithm is the balanced iterative reducing and clustering using hierarchies (BIRCH). Partition-based clustering methods start with each element as a single cluster and iteratively merge or reallocate them until a stopping criteria is met. Algorithms such as the K-Means or K-Medoids are partition-based methods. Density-based clustering methods examine the density of the data points and treat clusters as dense regions of the data space. A very popular algorithm is the density-based spatial clustering of applications with noise (DBSCAN). Grid-based clustering algorithms divides the data space into a finite number of cells before inspecting the density of data in each cell. Denser cells are connected to form clusters. Grid-based clustering algorithms are primarily developed for analyzing large spatial datasets. A well-known algorithm is the clustering in quest (CLIQUE).

A fundamental attribute of clustering algorithms is the similarity function (or its inverse, the distance function) that they use to assess data elements. Some studies \cite{toohey2015trajectory, Magdy2016613} review the most common similarity metrics used with trajectories. Among the popular metrics are the Euclidean distance, based on the spatial distance between trajectories; the Fr\'echet metric, based on the minimal spatial distance between any points in the two trajectories being compared; the dynamic time warping (DTW), where trajectories are undergo a non-linear transformation before comparison with another distance function; the longest common subsequence (LCSS), in which trajectories are stretched while other points remain unmatched in an attempt to provide an accurate similarity analysis; and the edit distance, which counts the minimum number of edits required to make the two trajectories equivalent.

Some researchers \cite{Furtado2016280} have investigated ways to calculate the similarity between the semantic trajectories, in which inferences about the nature of the trajectory are made. However, the research in this field is still in the initial stages and many challenges still need to be addressed.

\subsection{Big Data Processing Support}
    
Big Data relates to the phenomenon observed in the last decades, in which current digital technologies produce increasingly larger amounts of generated data, and as a consequence, changed the way data is captured, stored, analyzed, and visualized \cite{Mikalef20171}. This area is usually described in terms of three V's: Volume (amount of data), Variety (diverse structure of the data) and Velocity (the speed of data). Some of the well-known Big Data support tools include Spark, YARN, Hadoop, and MapReduce.

The trajectory domain also generates large amounts of data owning to the widespread use of GPS devices \cite{Janssens20131}. Therefore, new data capture, storage, processing, and visualization techniques are developed to handle the new reality in the domain.

 Big spatial-temporal frameworks have been developed to address some of these concerns. For example, the work in \cite{Alarabi201740} proposes an extension to the Hadoop framework for spatial-temporal data called ST-Hadoop, that includes contributions from indexing, to processing operations, and a SQL-like language to manipulate data. Sipresk \cite{Khazaei2016419} is a framework to perform analyses involving urban transportation data. It integrates spatial-temporal and trajectory data from many different sources and leverages traditional big data technologies to provide analytics. Lastly, Simba \cite{Xie2016} extends Spark to provide a fast, scalable solution with high throughput to data processing in the trajectory domain. The framework optimizes memory management and compares the results with other traditional spatial-temporal frameworks.

\section{State of the Art Challenges}
\label{State of the Art Challenges}

\paragraph*{The need for similarity metrics based on context and semantics} Trajectory data representation is described in terms of their most basic dimensions: spatial and temporal. For that reason, it is also referred to as raw trajectory. However, data representation can be improved with the addition of trajectory contextual and semantic dimensions. The context of a trajectory includes information such as the vehicle model, the driver, or other elements along the path created by the trajectory. Trajectory semantics are the reasons for the movement \cite{Shekhar2015799} (e.g. going home, or going to eat). Capturing new dimensions means having a richer trajectory description to analyze. As a consequence, new similarity metrics leverage the improved representation, creating new ways of comparing trajectories.

\paragraph*{The lack of spatial-temporal clustering techniques that leverage the improved trajectory similarity} Similarity definitions are usually based on a distance function between trajectories. Some trajectory similarity algorithms, such as the Dynamic Time Warping \cite{BLTJ:BLTJ272}, are based on and limited to only the spatial-temporal properties of each point of the trajectory. However, the relationship between two trajectories can be described in new forms, based on how close they are, and wether their paths overlap. As a consequence, important analysis conclusions may be missing, and clustering techniques are not expressive enough \cite{Adams2014118}.

\paragraph*{The need for an implemented big data spatial-temporal clustering framework} Since the trajectory domain generates  large amounts of data, hence big spatial-temporal data, any handling of this data becomes a challenging task. In addition, new trajectory data description and relationship modeling opens up space for new strategies in analysis. Several studies still point to the need for improvements in the integration of \cite{Le201258}, analysis of \cite{Kelley201441}, and infrastructure support for \cite{Adams2014118} trajectory data. This further demonstrates that a big data framework to process improved trajectory data in novel ways is still lacking.

\section{Spatial-Temporal Clustering Framework}
\label{Spatial-Temporal Clustering Framework}

Our work in progress is based on three objectives: \begin{enumerate*}[label=(\roman*)] \item to improve trajectory description with new contextual and semantic dimensions, so that novel forms of comparison can be calculated; \item to propose new trajectory clustering techniques that leverage new similarity modeling in trajectory data; and \item to present a framework that handles analysis in richer trajectory data. The activities associated with each of these goals are described in the following paragraphs.\end{enumerate*}

The first objective is to investigate current spatial-temporal data representation and to propose a richer trajectory data description. To accomplish this objective, ontologies and other studies in the literature are investigated. The main impact is on the trajectory data description. Our work currently investigates an augmented description of trajectory data with the addition of two dimensions: context and semantic. The context of a trajectory relates to the information that surrounds and relates to it, but are not spatial or temporal. An example is a vehicle on different weather conditions during a trip. The weather temperature is a dynamic contextual information that can be different at each timestamp of the trajectory.

The semantic of a trajectory relates to the reasons or goals that motivated an object to move. A simple example is a person who is going to work in the morning. More complex examples are going to lunch, or countourning a mountain. The semantics of a trajectory usually involves associating semantic attributes to segments of the trajectory itself.

The creation of new similarity relations is the second objective of this paper. The main activities to accomplish this objective involve exploring the expressiveness of the novel trajectory representations to define new similarity functions. It starts with an observation of the new trajectory description proposed in the previous goal, an investigation of how trajectory relationships can be described in terms of point to point relationships, and a comparison of small, point to point relationships to improve the overall trajectory comparison. For example, trajectories that share longer and continuous number of points (an overlap) may be scored higher than trajectories that share the same amount of points but scattered along their paths. As a consequence two pairs of trajectories that scored the same value by a similarity algorithm can be further analyzed and separated, based on new relationships. The main impact of this goal is on the trajectory clustering techniques. Clustering analyses will be richer, allowing novel inferences to be derived in the domain.

Lastly, the third objective addresses the development of a software framework to support novel and richer clustering techniques, and demonstrates the benefits of improved representation and similarity calculation.

\section{Conclusion}
\label{Conclusion}

This paper describes ongoing research towards an improved description of trajectory data, including contextual and semantic dimensions, novel similarity functions for trajectory clustering, and a framework to support the new forms of spatial-temporal clustering.

\bibliographystyle{IEEEtran}
\bibliography{../bibliography/scopus,../bibliography/noscopus}

\end{document}